\documentclass[prl,twocolumn,showpacs]{revtex4}

\usepackage{graphicx,epsfig,subfigure,amsmath,amssymb,amsfonts,euscript}
\usepackage{dcolumn}
\usepackage{bm}

\begin{document}

\newtheorem{assertion}{Assertion}


\title{Effects of interfacial curvature on Rayleigh-Taylor instability}

\author{Rouslan Krechetnikov}
\affiliation{University of Alberta, Edmonton, Canada}%

\date{April 19, 2008}

\begin{abstract}
In this work a non-trivial effect of the interfacial curvature on
the stability of accelerated interfaces, such as liquid rims, is
uncovered. The new stability analysis, based on operator and
boundary perturbation theories, reveals and quantifies influence
of the interfacial curvature on the growth rate and on the
wavenumber selection of the Rayleigh-Taylor instability. The
systematic approach developed here also provides a rigorous
generalization of the widely used \textit{ad hoc} idea, due to
Layzer [Astrophys. J. \textbf{122}, 1-12 (1955)], of approximating
the potential velocity field near the interface.
\end{abstract}

\maketitle

Accelerated interfaces \cite{Taylor:I,Rayleigh} are ubiquitous in
Nature and often exhibit long-wave Rayleigh-Taylor (RT)
instability, which occurs if the light fluid is accelerating into
the heavy one. The RT instability experienced by a liquid phase of
density $\rho$ and surface tension $\sigma$ can be described by
the time-evolution equation \cite{Taylor:I} for interfacial
perturbations $f(t)$, i.e. deviations from the flat interface, of
wavenumber $k$ under constant acceleration $g$ in the coordinate
system fixed in the interface
\begin{align}
\label{model_Taylor} {\mathrm{d}^{2} f(t) / \mathrm{d} t^{2}} = -
|k| \, \left[{\sigma \, \rho^{-1}} k^{2} + g\right] \, f(t).
\end{align}
Equation \eqref{model_Taylor} is given for the case when density
of one of the phases can be neglected, i.e. for unit Atwood
number. Apparently, if $g<0$ then the initially non-zero
perturbations will grow exponentially in time. The RT instability
\cite{Sharp} and its impulsive limit -- the Richtmyer-Meshkov
instability \cite{Richtmyer,Meshkov,Brouillette} independent of
the direction of acceleration -- are common in various phenomena:
e.g. combustion \cite{Khokhlov}, inertial confinement fusion
\cite{Lindl}, astrophysics \cite{Arnett,Arons,Cattaneo,Frieman},
geophysics \cite{Sazonov,Wilcock}, and many others. Because of
this wide fundamental impact, this classical instability still
attracts attention: in particular, there is a number of nonlinear
analyses \cite{Ott,Mikaelian,Hecht,Velikovich,Hazak} starting with
the seminal work of Layzer \cite{Layzer}, who proposed an
\textit{ad hoc} approximation of the velocity potential near the
tip of a finger leading to a nonlinear model for the finger
evolution. Despite numerous studies of the RT instability, the
influence of the interfacial curvature on its development has
never been pointed out in the literature. However, there are many
physical situations when curved interfaces are subject to
acceleration, e.g. in the drop splash problem \cite{Krechetnikov}.
Also thin liquid sheets with highly curved edges experiencing
accelerations are very frequent in various applications, but their
stability analysis is not yet available \cite{Sirignano}.

A nontrivial effect of non-zero interfacial curvature on the
stability characteristics can be understood in basic physical
terms. Namely, if we re-derive equation (\ref{model_Taylor}) for
the evolution of an interfacial disturbance of wavenumber $k$ for
flat interface using an energy argument, that is by evaluating
kinetic and potential energies of a perturbation, then it becomes
clear that the factor $|k|$ in (\ref{model_Taylor}) originates
from the fact that the perturbation penetrates in the bulk at the
distance $|k|^{-1}$. The latter is dictated by the solution of
Laplace's equation for the velocity potential $\phi \sim e^{|k| y
+ i k x}$ in a half-space, $(x,y) \in \Bbb{R} \times \Bbb{R}^{+}$.
In the case of a curved interface the penetration of a disturbance
into the bulk changes and thus the factor $|k|$ in
\eqref{model_Taylor} should be replaced with a function of both
the wavenumber $k$ and curvature, which clearly affects not only
the perturbation growth rate, but also the wavenumber selection!
However, formal stability analysis is more complicated than in the
flat interface base state case and requires accurate techniques to
solve for the velocity potential in a region with curved free
boundaries, as will be done below. The main results of physical
significance can be stated as follows: \textit{the interfacial
curvature and its sign influence the growth rate and the most
unstable wavenumber range selection of the RT instability.}

\begin{figure}
\centering
\epsfig{figure=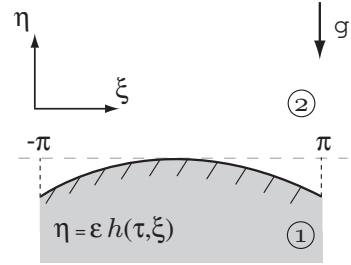,height=1.4in}
\caption{Curved interface as an
$O(\epsilon)$-perturbation.}\vspace{-0.85cm}
\label{sketch_interface_curved}
\end{figure}
In the analysis of the RT instability we adopt the Kelvin's
restrictive assumption \cite{Drazin}, i.e. consider an inviscid
and incompressible approximation of irrotational fluids. Let us
transform from the laboratory system $(x,y,t)$ to the one moving
with the interface with velocity $V(t)$ in the positive
$y$-direction: $\left(\xi=x,\eta=y-\int_{0}^{t}{V(\widetilde{t})
\, \mathrm{d}\widetilde{t}}, \tau=t\right)$ and $(u,v) \rightarrow
\left(\widetilde{u}=u,\widetilde{v}=v-V(t)\right)$, where tildas
stand for the variables in the moving frame of reference. Then the
potential function $\phi$, $\nabla \phi = (u,v)$, transforms into
$\widetilde{\phi} = \phi - V(t) \, \eta$, where we put the
arbitrary time-dependent constant of integration to zero without
loss of generality. Since $V(t)$ is not constant, in general, then
this new coordinate system is non-inertial and the full nonlinear
system for the bulk (the harmonic equation for the potential
$\widetilde{\phi}$ and the Lagrange-Cauchy integral for the
pressure $p$) and interfacial dynamics (the normal stress and
kinematic conditions) becomes
\begin{subequations}
\label{system_main_noninertial}
\begin{align}
\label{bulk_potential_NI}
\left\{\begin{array}{c} \Delta \widetilde{\phi} = 0, \\
\nabla \widetilde{\phi} \rightarrow 0, \ \eta \rightarrow -
\infty,
\end{array}\right., \ \eta \le \widetilde{f} \\
\label{bulk_pressure_NI} {\partial \widetilde{\phi} \over
\partial \tau} + {|\nabla \widetilde{\phi}|^{2} \over 2} = - {p \over \rho} -
\left(g + {\mathrm{d} V \over \mathrm{d} \tau}\right) \eta +
C(\tau), \ \eta \le \widetilde{f} \\
\label{interface_dynamic_NI} p = - {\sigma \widetilde{f}_{\xi\xi}
\over (1 + \widetilde{f}_{\xi}^{2})^{3/2}}, \ \eta = \widetilde{f} \\
\label{interface_kinematic_NI} {\partial \widetilde{f} \over
\partial \tau} + {\partial \widetilde{\phi} \over \partial \xi} {\partial \widetilde{f} \over
\partial \xi}= {\partial \widetilde{\phi} \over \partial \eta}, \ \eta =
\widetilde{f},
\end{align}
\end{subequations}
where $\nabla = \mathbf{i} \, \partial_{\xi} + \mathbf{j} \,
\partial_{\eta}$, $\widetilde{f}(\tau,\xi) = f(t,x) - \int_{0}^{t}{V(\widetilde{t}) \,
\mathrm{d}\widetilde{t}}$ is the position of the interface in new
coordinates. System \eqref{system_main_noninertial} is the
starting point of the stability analysis of curved interfaces; the
key idea is to consider the curved interface \textit{locally}, as
depicted in figure \ref{sketch_interface_curved}, with small
deviation from flatness, i.e. $\widetilde{f}(\tau,\xi) = \epsilon
h(\tau,\xi)$ with $\epsilon \ll 1$.

As one can infer from system \eqref{system_main_noninertial},
there exists a base state solution, which is motionless,
$\widetilde{\phi}^{0} = 0$, and steady, $\widetilde{f}^{0}(\xi)$,
if the conditions of static equilibrium are met, i.e.
\begin{align}
\label{base_state_curved_interface_RT} {\sigma \over
\rho}{\widetilde{f}_{\xi\xi} \over (1 +
\widetilde{f}_{\xi}^{2})^{3/2}} - \left(g + {\mathrm{d} V \over
\mathrm{d} \tau}\right) \widetilde{f} + C = 0,
\end{align}
where constant $C$ is time-independent. Obviously, surface tension
is required for the interface to have a non-zero curvature, while
the pure RT case ${\mathrm{d} V / \mathrm{d} \tau} = a =
\mathrm{const}$ allows the corresponding balance of the capillary
and hydrostatic pressures.

Next, it is important to construct a general solution of the
Laplace equation with non-fixed boundary values:
\begin{align}
\label{Laplace_curved_original}
\begin{array}{c}
\Delta \widetilde{\phi} = 0, \\
\eta = \epsilon h(\xi): \ \widetilde{\phi} =
\widetilde{\phi}_{0}(\xi),
\end{array}
\end{align}
where $\widetilde{\phi}_{0}(\xi)$ is an arbitrary summable
function to be determined from the free-boundary conditions. Here
we restrict the consideration to even functions $h(x)$ and to the
region in the neighborhood of the interface with the largest
curvature, as sketched in figure \ref{sketch_interface_curved}.
Without loss of generality, let the width of the domain be $2 \pi$
(in non-dimensional coordinates: \textit{cf}. figure
\ref{sketch_interface_curved}). Then we can construct the most
general velocity potential, which satisfies the Laplace equation
in this region and allows one to solve the free-boundary problem.
This constitutes the essence of the \textit{local approach}. Since
we are interested in a small, $O(\epsilon)$, perturbation of the
boundary (\textit{cf}. figure \ref{sketch_interface_curved}), then
it is natural to appeal to the boundary perturbation method. Its
basic idea \cite{Dyke} is to transform the boundary conditions on
the \textit{perturbed} boundaries to that on the
\textit{unperturbed} boundaries, which are known.

The main outcome of this analysis is that despite the fact that
the boundary is curved, as in figure
\ref{sketch_interface_curved}, finite Fourier modes, $e^{i n x}$,
constitute a complete set of functions and thus allow one to build
the solution to \eqref{Laplace_curved_original}, which can be
represented in the general form
\begin{align}
\label{potential_general_2D} \widetilde{\phi}(\tau,\xi,\eta) =
\sum_{n \in \Bbb{N}}{A_{n}(\tau) e^{|n| \eta} e^{i n \xi}}.
\end{align}
In this context, it is natural to comment on the \textit{ad hoc}
idea of Layzer \cite{Layzer}, which provided a break-through in
the nonlinear modelling of the RT and RM instabilities. Layzer
suggested approximating the potential function by
$\widetilde{\phi}(\tau,\xi,\eta) = A(\tau) e^{\eta} \cos{\xi}$
near the tip of the bubble, which, apparently, is just one
harmonic with $n=1$ out of the general expression
\eqref{potential_general_2D} and which allowed him to derive a
nonlinear evolution equation for the bubble amplitude $A(\tau)$.
Using more terms from \eqref{potential_general_2D} one can get a
more precise nonlinear model.

The equations for perturbations $\widetilde{f}'$ and
$\widetilde{\phi}'$ linearized around the base state curved
non-perturbed interface $\widetilde{f}^{0}(\xi)$ and
$\widetilde{\phi}_{0}$ in the frame moving with the interface are
given by:
\begin{subequations}
\label{system_2D_disturbances_NI_curved_II}
\begin{align}
{\partial \widetilde{\phi}' \over \partial \tau} &= - \left(g +
{\mathrm{d} V \over \mathrm{d} \tau}\right) \widetilde{f}' +
{\sigma \over \rho} \widetilde{f}'_{\xi\xi} + o(\epsilon), \\
{\partial \widetilde{f}' \over \partial \tau} &= - \epsilon
{\partial h^{0} \over \partial \xi} {\partial \widetilde{\phi}'
\over \partial \xi} + {\partial \widetilde{\phi}' \over \partial
\eta},
\end{align}
\end{subequations}
at $\eta=\epsilon h^{0}(\xi)$. Here we used the fact that
$\widetilde{f}^{0}(\xi) = \epsilon h^{0}(\xi)$, kept the terms up
to $O(\epsilon)$ and excluded pressure. As we will see, the term
of $O(\epsilon)$ will introduce a non-trivial correction to the
stability results for flat interfaces.

\begin{figure}
\centering \subfigure[Concave interface:
$\kappa>0$.]{\epsfig{figure=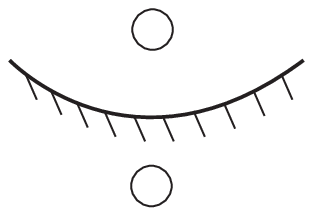,width=1.45in}\label{concave}}
\quad \subfigure[Convex interface: $\kappa<0$.
]{\epsfig{figure=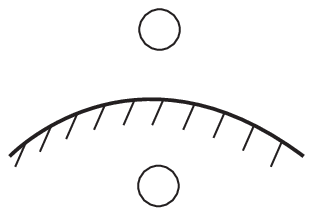,width=1.48in}\label{convex}}
\caption{Two generic curved interfaces; phase $1$ is the
(heaviest) liquid phase.}\vspace{-0.60cm}
\end{figure}

Since ${\mathrm{d} V / \mathrm{d} \tau} = a = \mathrm{const}$,
then system \eqref{system_2D_disturbances_NI_curved_II} contains
no explicit time-dependence and therefore one can perform the
standard eigenvalue analysis
$\left[\widetilde{\phi}'(\tau,\xi),\widetilde{f}'(\tau,\xi)\right]
\rightarrow \left[\Phi(\xi),F(\xi)\right] \, e^{\lambda \tau}$.
The subsequent analysis is based on the well-established operator
perturbation theory \cite{Kato}, which allows one to treat this
problem as a regular (non-singular) perturbation problem:
\begin{align}
\Phi(\xi) = \Phi^{0} + \epsilon \Phi^{1} + o(\epsilon), \ \lambda
= \lambda_{0} + \epsilon \lambda_{1} + o(\epsilon),
\end{align}
which yields \allowdisplaybreaks{
\begin{subequations}
\begin{align}
\label{eigenproblem_0} \epsilon^{0}&: & \lambda_{0}^{2} \Phi^{0} +
(g+a) \Phi^{0}_{\eta}
- {\sigma \over \rho} \Phi^{0}_{\xi\xi\eta} &= 0, \\
\label{eigenproblem_1} \epsilon^{1}&: & \lambda_{0}^{2} \Phi^{1} +
(g+a) \Phi^{1}_{\eta} - {\sigma \over \rho} \Phi^{1}_{\xi\xi\eta}
&= -  2
\lambda_{0} \lambda_{1} \Phi^{0} \\
& & + (g+a) h^{0}_{\xi} \Phi^{0}_{\xi} - {\sigma \over \rho}
\left[h^{0}_{\xi\xi\xi} \Phi^{0}_{\xi} \right. &\left.+ 2
h^{0}_{\xi\xi} \Phi^{0}_{\xi\xi}+ h^{0}_{\xi}
\Phi^{0}_{\xi\xi\xi}\right]. \nonumber
\end{align}
\end{subequations}}
\hspace{-0.2cm} Based on \eqref{potential_general_2D}, potentials
are given by
\begin{align}
\label{potential_function_approximations} \Phi^{i}(\xi,\eta) =
\sum_{n \in \Bbb{N}}{A_{n}^{i} e^{|n| \eta} e^{i n \xi}}.
\end{align}
Substituting the zero-order approximation $\Phi^{0}(\xi,\eta)$ in
\eqref{eigenproblem_0}, evaluating at $\eta=0$, and projecting
onto $e^{i n \xi}$ yields
\begin{align}
\label{eigenvalue_0} \lambda_{\pm 0}^{2} = - (g+a) |n| - (\sigma /
\rho) |n|^{3} = 0.
\end{align}
Next, substitution of $\Phi^{1}(\xi,\eta)$ into
\eqref{eigenproblem_1} and projection onto $e^{i n \xi}$ leads to
vanishing of the left-hand side of \eqref{eigenproblem_1} in view
of the definition of the zero-order eigenvalue
\eqref{eigenvalue_0}, while the rest of \eqref{eigenproblem_1}
results in
\begin{align}
&4 \pi \lambda_{0} \lambda_{1} A_{m}^{1} = \sum_{n}{A_{n}^{1}}
\int_{-\pi}^{\pi}\left\{i n (g+a) h^{0}_{\xi} \right.
\nonumber \\
&\left.- (\sigma / \rho) \left[i \, n \, h^{0}_{\xi\xi\xi} - 2
n^{2} h^{0}_{\xi\xi} - i \, n^{3} h^{0}_{\xi}\right]\right\}e^{i
(n-m) \xi}\mathrm{d}\xi. \nonumber
\end{align}
Since $\widetilde{\phi}'(\tau,\xi)$ is real, and thus
$\Phi^{i}(\xi,\eta)$ is real as well, then $A_{-r} = A_{r}$ and
thus expansions \eqref{potential_function_approximations} contain
only cosines. Since $h^{0}(\xi)$ and $h^{0}_{\xi\xi}(\xi)$ are
even functions of $\xi$, while $h^{0}_{\xi}(\xi)$ and
$h^{0}_{\xi\xi\xi}(\xi)$ are odd, then integrals involving
$h^{0}_{\xi}(\xi)$ and $h^{0}_{\xi\xi\xi}(\xi)$ should vanish. The
only terms left are
\begin{align}
4 \pi \lambda_{0} \lambda_{1} A_{m}^{1} = - 2 {\sigma \over \rho}
\sum_{n}{A_{n}^{1} \int_{-\pi}^{\pi}{n^{2} h^{0}_{\xi\xi} e^{i
(n-m) \xi}\mathrm{d}\xi}}. \nonumber
\end{align}
Since we perform the \textit{local analysis}, then $h^{0}_{\xi\xi}
\simeq \kappa = \mathrm{const}$ is the scaled $O(1)$ curvature and
the first approximation for the eigenvalue becomes
\begin{align}
\label{eigenvalue_1} \lambda_{\pm 1}^{(n)} \simeq {\sigma \over
\rho} {n^{2} \over \lambda_{\pm 0}^{(n)}}
\int_{-\pi}^{\pi}{h^{0}_{\xi\xi}\mathrm{d}\xi}.
\end{align}
Note that if the interface is non-symmetric, i.e. the odd
derivatives $h^{0}_{\xi}(\xi)$ and $h^{0}_{\xi\xi\xi}(\xi)$ do not
vanish, then these will affect the eigenvalue corrections; here we
consider the symmetric case, as the leading order effect. Hence,
we have proved the following
\begin{assertion}
\label{thm_2D_curved} If the flat interface is unstable in the RT
case, i.e. there exists real $\lambda_{+ 0}^{(n)}>0$, then the
addition of a positive curvature (concave interface: \textit{cf}
figure \ref{concave}) makes the physical system more unstable,
while the addition of negative curvature (convex interface:
\textit{cf} figure \ref{convex}) makes the system less unstable.
The eigenvalues obey
\begin{align}
\lambda_{\pm}^{(n)} = \lambda_{\pm 0}^{(n)} + \epsilon
\lambda_{\pm 1}^{(n)} + o(\epsilon),
\end{align}
with $\lambda_{\pm 0}^{(n)}$ and $\lambda_{\pm 1}^{(n)}$ given by
\eqref{eigenvalue_0} and \eqref{eigenvalue_1}, respectively.
\end{assertion}
In order to appreciate these results, let us make the following
two corollary type clarifications. First, the interpretation of
these curvature effects is not as trivial as one might expect,
i.e. that the presence of surface tension tends to flatten the
interface, since the curved interface base state is truly an
equilibrium base state. Second, because of the curvature effect,
the RT instability can be reversed, i.e. the sign of the growth
rate can change as a function of base state curvature! Indeed, if
the heavy phase $1$ accelerates the light phase $2$ and the
interface is flat, then there should be no instability according
to the RT criterion; however, if the interface is concave (cf.
figure \ref{concave}), then the instability may appear. In fact,
this can be illustrated with the well-known phenomena of
vapor-filled underwater collapsing bubbles
\cite{Birkhoff,Plesset}, which are unstable despite that the
denser liquid is accelerated towards the lighter vapor. Moreover,
as indicated above, this instability result is unaffected by
surface tension, which just allows for the existence of a base
state (spherical bubble, in this case). The latter problem has
been studied exactly because of its spherical symmetry, but to the
author's knowledge the conclusion that this is a particular case
of the more general effect of interfacial curvature has never been
established.

Lastly, it is known that the RT theory is valid only in the small
amplitude limit, but when the interfacial distortions become
significant the rate of their growth deviates significantly from
predicted one by the RT theory \cite{Lewis}. Apparently, one of
the sources of these deviations is due to finger formation and
thus non-zero curvature: the fingers can be considered, in a
quasi-static approximation, as a new base state which is subject
to perturbations. The latter will have growth rate different from
the case of a flat interface base state, as we just proved.

\begin{figure}
\centering \epsfig{figure=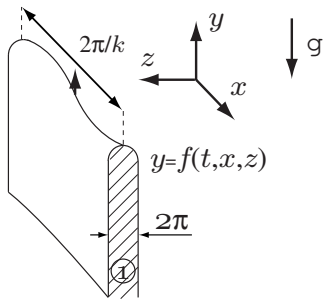,height=1.55in}
\caption{Three-dimensional rim.}
\label{3D_sketch_interface_curved}\vspace{-0.5cm}
\end{figure}

\begin{figure}
\centering \epsfig{figure=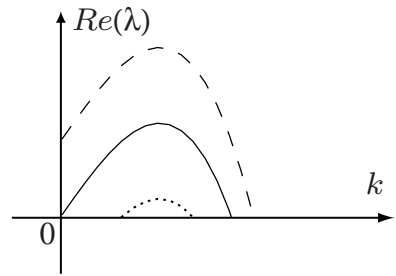,height=1.55in}
\caption{Effect of the interfacial curvature on the eigenvalues in
the 3D case. Solid curve corresponds to zero curvature, dashed
line to concave interface (positive curvature), and dotted line to
convex interface (negative curvature).}\vspace{-0.55cm}
\label{eigenvalues_plots}
\end{figure}

With the above understanding of the stability of two-dimensional
(2D) weakly curved interfaces, one can easily address the
stability of three-dimensional (3D) rims. Naturally, the main
question of interest is the rim instability along $x$-axis, as
shown schematically in figure \ref{3D_sketch_interface_curved}.
The idea is to analyze the structure of the solution near the rim
tip. Then, the translation of the previous results onto the 3D
case turns out to be straightforward, as suggested by the
structure of the velocity potential solution, i.e. the solution of
the 3D version of the problem \eqref{Laplace_curved_original}. The
zero- and first-order approximations read
\begin{align}
\label{potential_approximations_3D} \phi^{(j)}(x,y,z) = \sum_{n
\in \Bbb{N}}{A_{n}^{(j)} e^{i k x} e^{\sqrt{k^{2}+|n|^{2}} y} e^{i
n z}},
\end{align}
with $j =1,2$ and therefore one gets the eigenvalue approximations
$\lambda_{\pm 0}^{(n)}$ and $\lambda_{\pm 1}^{(n)}$ analogous to
\eqref{eigenvalue_0} and \eqref{eigenvalue_1}:
\begin{align}
\lambda_{\pm 0}^{(n)} &= \pm \left[- (g+a) \sqrt{k^{2}+|n|^{2}} -
{\sigma \over \rho} \left(k^{2}+|n|^{2}\right)^{3/2}\right]^{1/2},
\nonumber \\
\lambda_{\pm 1}^{(n)} &\simeq {\sigma \over \rho} {n^{2} \over
\lambda_{\pm 0}^{(n)}}
\int_{-\pi}^{\pi}{h^{0}_{\xi\xi}\mathrm{d}\xi}, \nonumber
\end{align}
i.e. the only difference is that the discrete wavenumber, $n$, (in
$z$-direction) in \eqref{eigenvalue_0} is replaced with the
wavenumber in the $(x,z)$-plane, $\sqrt{k^{2}+|n|^{2}}$. This fact
and equation \eqref{potential_approximations_3D} for the
perturbation velocity field allow one to clearly see the
previously made energy argument: the curvature affects the depth
of penetration of a disturbance into the bulk and thus the factor
$|k|$ in (\ref{model_Taylor}) is modified. In fact, for the
long-wave perturbations of a rim of a liquid sheet of thickness $2
\pi$, i.e. when $|k| \gg (2 \pi)^{-1}$, the depth of penetration
is $\sim 2 \pi$ and therefore the factor $|k|$ in
(\ref{model_Taylor}) is replaced by $(2 \pi)^{-1}$. The latter of
course changes the growth rate and the wavenumber selection.
Since, the eigenvalue has the following structure:
\begin{align}
\lambda = \lambda_{0}(k;n) + C(n) \kappa,
\end{align}
where $\kappa$ is the curvature, then \textit{the curvature in
$z$-direction has an effect on the wavenumber selection in
$x$-direction}, as one can learn from figure
\ref{eigenvalues_plots}. Namely, concave interfaces enlarge the
range of unstable wavelength for fixed surface tension (dashed
line in figure \ref{eigenvalues_plots}), while convex interfaces,
as in figure \ref{3D_sketch_interface_curved}, stabilize the
physical system and thus narrow the range of unstable wavenumbers
(dotted line in figure \ref{eigenvalues_plots}). Thus, the
analysis of 3D curved interfaces can be summarized as follows
\begin{assertion}
\label{thm_3D} The stability of 3D rims, as that shown in figure
\ref{3D_sketch_interface_curved}, is affected by the transverse
curvature: concave interfaces are less stable than flat ones,
while convex interfaces are more stable. The range of
lengthwise-unstable wavenumbers (i.e. along-the-rim wavenumbers)
is narrowed in the case of convex interfaces and widened in the
case of concave interfaces.
\end{assertion}

In conclusion, the main contribution of this study is the
clarification of the interfacial curvature effects in the 2D and
3D cases on the growth rate and the wavenumber selection of the
Rayleigh-Taylor instability. All the major results are summarized
in Assertions \ref{thm_2D_curved}-\ref{thm_3D}, and can be easily
extended onto the case of Richtmyer-Meshkov instability. The
analysis of the stability of curved interfaces also leads to the
rigorous generalization of the classical idea due to \cite{Layzer}
on approximating the potential function in free-boundary problems
with curved base state interfaces. While the base state
interfacial curvature considered in this Letter is due to the
presence of surface tension (for the sake of mathematical
concreteness), it should have analogous implications for the
stability characteristics regardless of its physical origins.

\begin{acknowledgements}
The author gratefully acknowledges valuable feedback and
encouraging discussions with Professor Bud Homsy.
\end{acknowledgements}

\end{document}